\documentclass[fleqn,12pt,twoside]{article}
\usepackage{espcrc1}
\usepackage{epsfig}  

\usepackage{graphicx}
\usepackage[figuresright]{rotating}

\newcommand{\AmS}{{\protect\the\textfont2
  A\kern-.1667em\lower.5ex\hbox{M}\kern-.125emS}}

\title{(di)lepton physics with ALICE}

\author{P.~Crochet\address[]{Laboratoire de Physique Corpusculaire\\
    CNRS/IN2P3 and Universit\'e Blaise Pascal\\
    F-63000 Clermont-Ferrand, France}$^{\rm ,}$\thanks{crochet@clermont.in2p3.fr}
    for the ALICE Collaboration}
       
\begin{document}

\maketitle

\begin{abstract}
Physics perspectives with (di)lepton measurements with the ALICE detector
at the LHC are reviewed. Special emphasis is placed on heavy flavor physics.
\end{abstract}

\section{(di)leptons and heavy flavors: what is different at the LHC}
One of the most exciting aspects of heavy ion collisions at the LHC is 
the abundant production rate of heavy flavors which can be used, for the 
first time, as high statistics probes of the 
medium~\cite{Carminati:2004fp,Bedjidian:2003gd}.
This allows to use a large variety of new observables.
The magnitude of most of the in-medium effects is expected to be dramatically 
enhanced. 
Some of these aspects are discussed hereafter.

\begin{itemize}
\item{{\bf Large primary production cross-sections:}
The number of $c\bar{c}$ ($b\bar{b}$) pairs produced in central $AA$ 
collisions at the LHC is expected to be 10 (100) times larger than at RHIC.
Therefore, at the LHC both charmonia and bottomonia can be used, thus 
providing powerful probes for Quark Gluon Plasma (QGP) studies.
In fact, since the $\Upsilon(1S)$ state is expected to only dissolve
significantly above the 
critical temperature~\cite{Digal:2001ue,Wong:2004zr},
which might only be reachable at the LHC, the spectroscopy of the 
$\Upsilon$ family should reveal unique characteristics of the 
QGP~\cite{Gunion:1996qc};}
\item{{\bf Large resonance dissociation rate:}
In addition to nuclear absorption, comoving hadrons and color screening, 
quarkonia can be significantly destroyed by gluon 
bombardment~\cite{Xu:1995eb}.
This mechanism which results from the presence of quasifree gluons starts 
being effective for temperatures above the critical temperature but 
not necessarily above the resonance dissociation temperature
by color screening.
It is expected to be relatively important at the LHC.
Indeed, recent estimates~\cite{Bedjidian:2003gd} 
of the dissociation cross-sections show that none of 
the prompt $J/\psi$ would survive the deconfined phase at the LHC
and that about 80\% of the $\Upsilon$ would be destroyed;}
\item{{\bf Large charmonium secondary production:}
Besides indirect charmonia production from $b$-hadron 
decay~\cite{Eidelman:2004wy} (see below) an important yield of 
secondary charmonia is expected from $D\bar{D}$ annihilation~\cite{Ko:1998fs},
statistical hadronization~\cite{Braun-Munzinger:2000px}
and kinetic recombination~\cite{Thews:2000rj}.
The two last processes explicitly assume the formation of a deconfined medium.
The underlying picture is that charmonium resonances form by coalescence of 
free $c$ and $\bar{c}$ in the QGP~\cite{Thews:2000rj} or at the 
hadronization stage~\cite{Braun-Munzinger:2000px}.
According to these models, the signature of the QGP should lead to an 
increase of the $J/\psi$ yield versus centrality, proportional to 
${\rm N}^2_{c\bar{c}}$, instead of a suppression\footnote{Note that, due
to the large number of $c\bar{c}$ pairs produced in central heavy ion 
collisions at LHC, these models predict a spectacular enhancement of 
the $J/\psi$ yield, up to a factor 100 in central collisions, relative to the 
primary production yield~\cite{Bedjidian:2003gd,Andronic:2003zv}.};}
\item{{\bf Complex structure of (di)lepton spectra:}
With a low $p_{\rm t}$ threshold of about 2~GeV/c on the decay leptons, 
unlike-sign dileptons from bottom decay dominate the dilepton correlated 
component all over the mass range.
Whereas in the high invariant mass region each lepton comes from the direct 
decay of a $B$ meson, in the low invariant mass region both leptons 
come from the decay of a single $B$ meson via a $D$ meson.
Next-to-leading order processes such as gluon splitting also populate 
significantly the low mass dilepton spectrum due to their particular 
kinematics~\cite{Norrbin:2000zc}.
Then, a sizeable yield of like-sign correlated dileptons from bottom 
decay is present. 
This contribution arises from the peculiar decay chain of $b$ hadrons and 
from $B$-meson oscillations.
The single lepton spectra are also subject to significant novelties. 
The most striking one is the emergence of the $W^\pm$ bosons as a bump 
located at around $30~{\rm GeV/c}$ in the single lepton $p_{\rm t}$ 
distributions~\cite{zaida}.}
\end{itemize}


\section{Selected physics channels}
ALICE (A Large Ion Collider Experiment) is the LHC experiment
dedicated to the study of nucleus-nucleus collisions.
The detector consists of a central barrel ($|\eta|<0.9$), a forward muon
spectrometer $(2.5<\eta<4$) and several forward/backward and central small 
acceptance detectors~\cite{Carminati:2004fp,Hans-Ake}.
(di)leptons will be measured in ALICE through the electron channel in the
central region and through the muon channel in the forward region.
Selected physics channels are presented below.

\begin{itemize}
\item{{\bf $\Upsilon^\prime/\Upsilon$ ratio versus $p_{\rm t}$:} 
The $p_{\rm t}$ suppression pattern of a resonance is a consequence of
the competition between the resonance formation time and the QGP 
temperature, lifetime and spatial extent~\cite{Blaizot:1987ha}.
Quarkonium suppression is expected also as the result of nuclear effects 
like shadowing and absorption.
In order to isolate pure QGP effects, it has been proposed to study the 
$p_{\rm t}$ dependence of quarkonium ratios instead of single quarkonium 
$p_{\rm t}$ distributions.
By doing so, nuclear effects cancel out, at least in the $p_{\rm t}$ 
variation of the ratio.
Following the arguments of~\cite{Gunion:1996qc}, 
the capabilities of the ALICE muon spectrometer to measure the 
$p_{\rm t}$ dependence of the $\Upsilon^\prime/\Upsilon$ ratio in central 
(10\%) Pb-Pb collisions have been investigated~\cite{ericTHESIS}.
Two different QGP models with different system sizes were considered.
The results of the simulations show that, with the 
statistics collected in one month of data taking, the measured 
$\Upsilon^\prime/\Upsilon$ ratio 
exhibits a strong sensitivity to the characteristics of the QGP;}
\item{{\bf Secondary $J/\psi$ from $b$-hadron decay:} 
As stated above, a large fraction of $J/\psi$ arises from $b$-hadron 
decay~\footnote{In central (5\%) Pb+Pb collisions at 
$\sqrt{s} = 5.5~{\rm TeV}$, 
${\rm N}(b\bar{b}\rightarrow J/\psi)/{\rm N}({\rm direct}~J/\psi) \sim 20\%$
in $4\pi$ (with shadowing and feed-down but without nuclear 
absorption)~\cite{Bedjidian:2003gd}.}. 
These secondary $J/\psi$, which are not QGP suppressed, must be subtracted 
from the measured $J/\psi$ yield prior to $J/\psi$ suppression studies.
They can be identified by exploiting the large lifetime of $b$ hadrons 
which results in a finite impact parameter for the decay leptons of secondary 
$J/\psi$.
Simulations have shown that such measurements can successfully be performed
with dielectrons measured in the central part of ALICE 
thanks to the excellent spatial resolution of the Inner Tracking 
System~\cite{TRDTP};}

\item{{\bf Open heavy flavors:} 
The open heavy flavor cross-section can be measured by means of several 
channels: low-mass and high-mass unlike-sign dileptons~\cite{Rachid}, 
single lepton $p_{\rm t}$ distributions~\cite{Rachid,Padova}, 
like-sign dileptons~\cite{Crochet:2001qd}, single leptons with displaced 
vertices~\cite{TRDTP,Padova}, 
secondary $J/\psi$ from $b$-hadron decay~\cite{TRDTP} and electron-muon
coincidences~\cite{TRDTP}.
Recently, a measurement of the differential inclusive $b$-hadron cross-section 
has been investigated in the electron channel~\cite{Padova} with a technique 
developed in $p\bar{p}$ collisions~\cite{Albajar:1988th} 
and adapted to heavy ion collisions in the ALICE-muon channel~\cite{Rachid}.
The results presented in Figure~\ref{fig} (left) show that the $b$-hadron 
cross-section
can be reconstructed up to $p_{\rm t}^{b\;{\rm hadron}} = 30~{\rm GeV/c}$.
Sensitivity to the $b$-quark energy loss is evidenced such that
the nuclear modification factors, which can be simultaneously measured for
light hadrons, for $D^0$~\cite{Dainese:2003wq} and for $b$ hadrons should 
provide a set of powerful tools to investigate the mass dependence of 
the energy loss;}
\vspace*{-0.4cm}
\begin{figure}[hbt]
  \centering{\epsfig{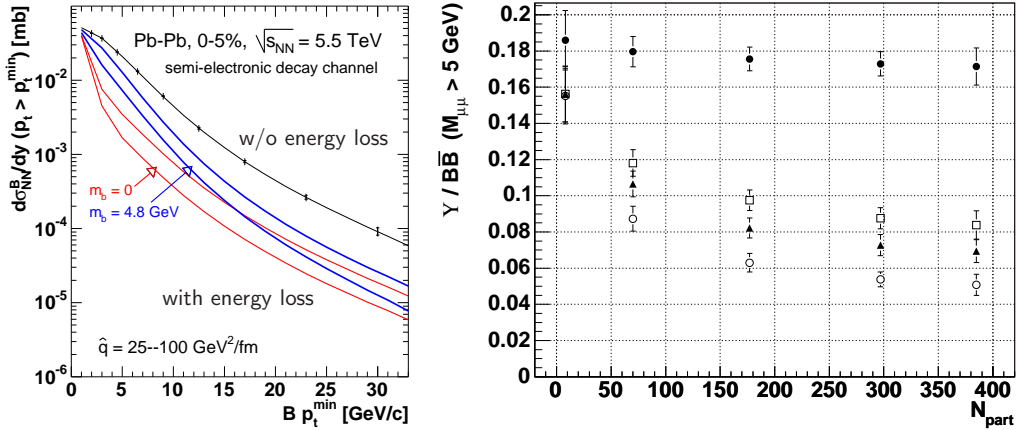}}
\vspace*{-0.8cm}
  \caption{Left: differential inclusive $b$-hadron cross-section reconstructed 
    from single electrons with displaced vertices
    in central (5\%) Pb+Pb collisions~\cite{Padova,PPRV2}.
    The results are shown without and with $b$-quark energy loss according 
    to~\cite{Armesto:2005iq}.
    Right: centrality dependence of the $\Upsilon/b\bar{b}$ ratio in Pb+Pb 
    collisions in the muon channel. The ratio is shown without (dots) 
    and with (squares) $\Upsilon$ 
    nuclear absorption as well as with nuclear absorption and 
    melting by color screening (triangles and circles) with dissociation 
    temperatures taken from~\cite{Wong:2004zr}.
    Taken from~\cite{PPRV2,SmbatPriv}.}
  \label{fig}
\end{figure}
\vspace*{-0.9cm}
\item{{\bf Centrality dependence of the $\Upsilon/b\bar{b}$ ratio:} 
If the $b$-quark energy loss turns out to be negligible, the $b$-hadron 
cross-section can be used as a normalisation for $\Upsilon$ suppression
studies.
This normalisation is the most natural normalisation because of 
the similar production processes for open and hidden heavy flavors.
Figure~\ref{fig} (right) shows the measurement that can be achieved 
in one month of Pb beams.
The triangles and circles illustrate the typical sensitivity
of the ratio to $\Upsilon$ melting by color screening.
Note that the statistical uncertainty of the ratio is dominated by the 
statistics of the probe (i.e. the number of $\Upsilon$s) and not 
by the statistics of the reference.
Indeed, the number of correlated unlike-sign muon pairs from bottom decay
in the mass range ${\rm M}_{\mu\mu} > 5~{\rm GeV/c}^2$ is larger than that 
of the $\Upsilon$ by a factor 5.}
\end{itemize}
\section{Summary}
(di)lepton measurements with the ALICE detector at the LHC will bring an 
unprecedentedly rich physics program in the heavy flavor sector of heavy ion 
collisions.
In addition to the channels discussed here, further exciting possibilities 
should be opened with, for example, quarkonia polarization and dilepton 
correlations.
\section*{Acknowledgments}
Part of this work was supported by the EU Integrated Infrastructure
Initiative HadronPhysics Project under contract number
RII3-CT-2004-506078. 


\end{document}